\documentclass[11pt,a4paper]{article}

\pdfoutput=1

\usepackage[UKenglish]{babel}
\usepackage[UKenglish,cleanlook]{isodate}
\usepackage{amsmath}
\usepackage{amssymb}
\usepackage[retainorgcmds]{IEEEtrantools}
\usepackage{dsfont}
\usepackage[square,comma,numbers,sort&compress]{natbib}
\usepackage{titling}
\usepackage{authblk}
\usepackage{tocloft}
\usepackage[colorlinks,linkcolor=blue,citecolor=blue,urlcolor=blue]{hyperref}
\usepackage{quant_defs}

\pretitle{\begin{center}\Large\sffamily\bf}
\posttitle{\par\end{center}}

\newcommand{\vect}[1]{\boldsymbol{#1}}

\newcommand{\A}[0]{\mathrm{A}}
\newcommand{\B}[0]{\mathrm{B}}

\newcommand{\E}[0]{\mathrm{E}}
\newcommand{\hilb}[0]{\mathcal{H}}
\newcommand{\ee}[0]{\mathrm{e}} 
\newcommand{\ii}[0]{\mathrm{i}} 
\newcommand{\etal}[0]{\textit{et al.}}

\newcommand{\work}{paper}

\newcommand{\Hc}[0]{H} 
\newcommand{\Hq}[0]{H} 

\title{A quantum cloning bound and application to quantum key distribution}
\author{Erik Woodhead\thanks{Erik.Woodhead@ulb.ac.be}}
\date{14 August 2013}
\affil{Laboratoire d'Information Quantique, CP 225, Universit\'e{} Libre de
  Bruxelles (ULB), Boulevard du Triomphe, 1050 Brussels, Belgium}

\begin{document}

\maketitle

\begin{abstract}
  We introduce a quantum cloning bound which we apply to a straightforward
  and relatively direct security proof of the prepare-and-measure
  Bennett-Brassard 1984 (BB84) quantum key distribution (QKD) protocol
  against collective attacks. The approach we propose is able to handle the
  practical problem of source and detector alignment imprecisions in a simple
  way. Specifically, we derive a keyrate bound for a BB84 implementation in
  which Alice's source emits four given but arbitrary pure states, where the
  usual equivalence between prepare-and-measure and entanglement-based QKD no
  longer applies. Our result is similar to a keyrate derived by Mar\o{}y
  \etal{} [\href{http://dx.doi.org/10.1103/PhysRevA.82.032337}{Phys.~Rev.~A
    \textbf{82},\ 032337 (2010)}] and generally an improvement over the
  keyrate derivable from the entropic uncertainty relation in situations
  where it applies. We also provide a stronger result for a source emitting
  arbitrary qubit states, and a further improved result if the detector is
  additionally assumed two dimensional.
\end{abstract}

\clearpage


\section{Introduction\label{sec:introduction}}

\subsection{Background and motivation}

Quantum key distribution (QKD) \cite{ref:bb1984} is an approach to the
problem of generating a cryptographic key and disseminating it securely
between its intended recipients. The general advantage of QKD is that it
permits the security of a key to be certified based only on the laws of
physics, in particular limitations inherent to quantum physics, rather than
assumptions about the computational power of a potential adversary.

There are two related classes of QKD implementation considered in the
literature: the entanglement-based approach \cite{ref:e1991}, in which the
secret key is generated by joint measurements on a shared entangled state,
and ``prepare-and-measure'' QKD \cite{ref:bb1984}, in which one party
(``Alice'') prepares states and transmits them to another (``Bob''). A simple
well-known argument \cite{ref:bbm1992} establishes a correspondence between
entanglement-based protocols and a restricted class of prepare-and-measure
protocols in which the source states satisfy a condition of basis
independence \cite{ref:kp2003} (where the average of the source states is the
same in the different bases used).

While this argument has played an important role in the development of
security proofs, it also has its limitations. Real world QKD implementations
are subject to various practical issues, including imperfections in the
devices, which can lead to a false certification of security if not accounted
for in the security analysis \cite{ref:wp2013}. In particular, most
experimental and commercially available QKD implementations are of the
prepare-and-measure variety. In this case the states used in the
implementation will not form orthogonal bases and will inevitably deviate
from the basis-independence condition due to the limited accuracy of the
source. Consequently, security proofs of entanglement-based QKD protocols
accounting for arbitrary detector alignment imprecisions -- such as the
recent security proofs of the entanglement-based Bennett-Brassard (BB84) BB84
protocol based on entropic uncertainty relations
\cite{ref:bc2010,ref:tr2011,ref:tl2012} -- do not translate to security
proofs accounting for all possible source alignment flaws in the
prepare-and-measure setting.

A few security proofs of the prepare-and-measure BB84 protocol have tackled
this issue based on a reduction to an equivalent virtual protocol
\cite{ref:gl2004,ref:k2009,ref:mls2010}. In this virtual protocol, Alice's
bit and basis choices are effectively both determined by the outcome of a
single four-outcome measurement on an entangled ``coin'' state that must be
partially or fully characterised and prepared by Alice. Conceptually, this
differs from security analyses of genuine entanglement-based QKD protocols,
which assume the entangled state is prepared by an adversary and unknown to
the legitimate parties apart from what they can deduce about it via their
measurement results.

In this \work{}, we propose an alternative approach to this problem in which
we abandon the use of entangled states altogether. We show that a
straightforward and relatively direct asymptotic security proof of the
prepare-and-measure BB84 protocol against collective attacks is possible
based on an appropriate characterisation of the limitations of quantum
cloning. The no-cloning theorem \cite{ref:wz1982,ref:d1982}, which implies
that an eavesdropper cannot copy non-orthogonal states without disturbing
them in a visible way, is often considered the intuitive basis for the
security of QKD. This intuitive connection was noted in the original proposal
of Bennett and Brassard \cite{ref:bb1984}, and is further discussed in
\cite{ref:sig2005}. The approach of studying the prepare-and-measure BB84
protocol ``directly'' was followed in some early generic security analyses
\cite{ref:m2001,ref:bb2002}, including the original security proof of BB84
against individual attacks \cite{ref:fg1997}, though has seen little
development following Shor and Preskill's security proof based on
entanglement distillation \cite{ref:sp2000}.

As our main result, we derive a secret keyrate for a BB84 implementation in
which the source emits four given but arbitrary pure states and which is
largely device independent on Bob's side. Our end result is comparable to the
keyrate derived by Mar\o{}y \etal{} \cite{ref:mls2010} which, to our
knowledge, is the only previous analysis that establishes security under such
relaxed conditions. Specifically, we recover the same keyrate as Mar\o{}y
\etal{} if the two $z$-basis source states are emitted with equal
probability, but show that it can be improved in the case of a biased source.

While it is interesting to have a security result that holds for arbitrary
source states, there are situations of practical interest where these states
are naturally qubits. This is the case if, for instance, Alice encodes her
basis and bit choices in the form of different photon polarisation states. In
such an implementation, the source states may still span a Hilbert space of
more than two dimensions in practice if, for example, an attenuated laser is
used to produce low intensity coherent states. Anticipating that this
complication is probably best addressed with the decoy-state method
\cite{ref:h2003,ref:lmc2005}, as a second main result we also present an
improved keyrate bound for a source emitting arbitrary qubit states. If we
additionally assume Bob's detector measurements are two dimensional, we find
that we can further improve our result to the point that, if no errors are
observed, we obtain an asymptotic keyrate of 1.

\subsection{Outline of scenario and main results}

We imagine Alice possesses a source which, depending on a choice of bit and
``basis'', emits one of four different quantum states which is sent to
Bob. Throughout this \work{} we will consistently call the two ``$z$-basis''
states $\ket{\alpha}$ and $\ket{\alpha'}$, and the two ``$x$-basis'' states
$\ket{\beta}$ and $\ket{\beta'}$ (note that this nomenclature is not intended
to imply that these states are necessarily really the $\sz$ and $\sx$
eigenstates). We denote the corresponding density operators by $\rho =
\proj{\alpha}$, $\rho' = \proj{\alpha'}$, $\sigma = \proj{\beta}$, and
$\sigma' = \proj{\beta'}$. Following a unitary attack by a potential
eavesdropper (``Eve''), these states are contained in some Hilbert space
$\hilb_{\A} \subset \hilb_{\B} \otimes \hilb_{\E}$, where $\hilb_{\B}$ and
$\hilb_{\E}$ respectively denote the Hilbert spaces accessible to Bob and
Eve. Upon reception of the states emitted by Alice, Bob performs one of two
binary-outcome measurements (a ``$z$-basis'' or ``$x$-basis'' measurement,
though we do not assume Bob is necessarily measuring $\sz$ or $\sx$) in order
to compute his version of the raw key. Alice and Bob then sacrifice a subset
of their results in order to estimate the $z$- and $x$-basis bit error rates
$\delta_{z}$ and $\delta_{x}$, and if these are not too high, extract a
secret key by error correction and privacy amplification.

We consider the common variant of BB84 where only the $z$-basis results are
used to generate the key. In this case, for an unbiased source an application
of the Devetak-Winter bound \cite{ref:dw2005} yields a lower bound on the
asymptotic secret keyrate $r$ which depends only on the fidelity
$F(\rho_{\E}, \rho'_{\E})$ between the parts of the $z$-basis states
accessible to the eavesdropper:
\begin{equation}
  \label{eq:keyrate_fidel}
  r \geq 1
  - h \bigl( \tfrac{1}{2} + \tfrac{1}{2} F(\rho_{\E}, \rho'_{\E}) \bigr)
  - h(\delta_{z}) \,.
\end{equation}
In this expression, the fidelity is defined by $F(\rho, \sigma) =
\trnorm{\sqrt{\rho}\sqrt{\sigma}}$ where $\trnorm{\sps{\cdot}}$ denotes the
trace norm, $h(x) = - x \log(x) - (1 - x) \log(1 - x)$ is the binary entropy,
all logarithms are in base 2, and subscripts indicate partial tracing in the
usual way (e.g.\ $\rho_{\E} = \Tr_{\B}[\rho]$). Bounding the keyrate is then
reduced to lower bounding $F(\rho_{\E}, \rho'_{\E})$. Alice and Bob, by
contrast, may lower bound the trace distances $D(\rho_{\B}, \rho'_{\B})$ and
$D(\sigma_{\B}, \sigma'_{\B})$ in terms of the observed bit error rates via
the Helstrom bound \cite{ref:h1976}: $D(\rho_{\B}, \rho'_{\B}) \geq \abs{1 -
  2\delta_{z}}$ and $D(\sigma_{\B}, \sigma'_{\B}) \geq \abs{1 -
  2\delta_{x}}$, where we define the trace distance by $D(\rho, \sigma) =
\tfrac{1}{2} \trnorm{\rho - \sigma}$. Our derivation is then completed by
bounding $F(\rho_{\E}, \rho'_{\E})$ in terms of $D(\sigma_{\B},
\sigma'_{\B})$. Any such bound can be understood as a fundamental limit on
the degree to which quantum states can be cloned.

In general, such a bound will depend on a characterisation of the states
$\ket{\alpha}$, $\ket{\alpha'}$, $\ket{\beta}$, and $\ket{\beta'}$ emitted by
the source. Similar to the authors of \cite{ref:gl2004,ref:mls2010}, we will
consider the case where we are given a ``basis overlap'' angle $\theta$,
which we define such that
\begin{equation}
  \label{eq:theta_def}
  \sqrt{1 + \abs{\sin(\theta)}}
  = \tfrac{1}{2} \babs{\braket{\alpha}{\beta} +
    \braket{\alpha'}{\beta} + \braket{\alpha}{\beta'} -
    \braket{\alpha'}{\beta'}} 
\end{equation}
wherever the right-hand side is greater than 1. Our main cloning bound then
reads
\begin{equation}
  \label{eq:fidel_theta_bound}
  F(\rho_{\E}, \rho'_{\E})
  \geq f_{\theta} \bigl( D(\sigma_{\B}, \sigma'_{\B}) \bigr) \,,
\end{equation}
with the function $f_{\theta}$ defined by
\begin{equation}
  \label{eq:def_f_theta}
  f_{\theta}(v)
  = \begin{cases}
    \abs{\sin(\theta)} v - \abs{\cos(\theta)} \sqrt{1 - v^{2}}
    &: v \geq \abs{\cos(\theta)} \\
    0 &: v \leq \abs{\cos(\theta)}
  \end{cases}\,,
\end{equation}
which provides a lower bound on $F(\rho_{\E}, \rho'_{\E})$ wherever the
right-hand side of \eqref{eq:theta_def} is greater than 1.

These are the essential ingredients of our security analysis. Bounding the
keyrate is a matter of applying \eqref{eq:fidel_theta_bound} and
$D(\sigma_{\B}, \sigma'_{\B}) \geq \abs{1 - 2\delta_{x}}$ to
\eqref{eq:keyrate_fidel}. The resulting keyrate is then
\begin{equation}
  \label{eq:keyrate_theta}
  r \geq 1 - h \bigl( \tfrac{1}{2}
  + \tfrac{1}{2} f_{\theta}( \abs{1 - 2\delta_{x}} ) \bigr)
  - h(\delta_{z}) \,.
\end{equation}
This coincides with the keyrate given in \cite{ref:mls2010} for the scenario
we have outlined here, though as will be discussed later on, our general
approach can lead to improved keyrates in certain cases, specifically where
the source is biased or is limited to two dimensions and more information
about the states is used.

We give a derivation of the asymptotic keyrate bound \eqref{eq:keyrate_fidel}
for an unbiased source in section~\ref{sec:bounding_keyrate}, as well as a
generalisation for a biased source where the two $z$-basis states are not
emitted with equal probability. We also briefly comment on the issue of
non-asymptotic security, which we mostly leave as a problem for future
work. In section~\ref{sec:fidel_bounds}, we discuss how an eavesdropper's
distinguishing power between the two $z$-basis states, measured in terms of
the fidelity as described above, can be bounded. This includes a proof of the
cloning bound \eqref{eq:fidel_theta_bound} introduced above, a better bound
applicable if the source states are qubits, and a further improved bound if
both the source and detector are two dimensional. We finish with some
concluding remarks and summarise potential directions for future research in
section~\ref{sec:conclusion}.

\section{Bounding the keyrate\label{sec:bounding_keyrate}}

\subsection{Derivation of asymptotic keyrate bound}

Assuming an eavesdropper restricted to collective attacks, a lower bound on
the asymptotic secret keyrate that can be securely extracted from Alice's
sifted key by one-way error correction and privacy amplification is given by
the Devetak-Winter bound \cite{ref:dw2005}:
\begin{equation}
  \label{eq:devetak_winter}
  r \geq \Hq(Z \mid \E) - \Hc(Z \mid Z') \,.
\end{equation}
Here, the conditional Shannon entropy $\Hc(Z \mid Z')$ quantifies the key
loss due to error correction, and is upper bounded by $h(\delta_{z})$ (with
equality in the typical case of symmetric errors). The conditional von
Neumann entropy $\Hq(Z \mid \E)$ quantifies the amount of key that can be
extracted securely by privacy amplification, which for our purposes we should
evaluate on the classical-quantum state
\begin{equation}
  \label{eq:cq_ae}
  \tau_{Z\E} = \tfrac{1}{2} \bigl( \proj{0}_{Z} \otimes \rho_{\E}
  + \proj{1}_{Z} \otimes \rho'_{\E} \bigr) \,,
\end{equation}
where $\rho$ and $\rho'$ are the $z$-basis states emitted by Alice, and
$\ket{0}_{Z}$ and $\ket{1}_{Z}$ denote the state of a classical register in
her possession. The state \eqref{eq:cq_ae} expresses the correlation between
Alice's version of the sifted key and the corresponding quantum states in
Eve's possession.

In order to arrive at \eqref{eq:keyrate_fidel}, we use that $\Hq(Z \mid \E)
\geq \Hq(Z \mid \E\E')$ for any extension of the state \eqref{eq:cq_ae} to a
larger Hilbert space $\hilb_{Z} \otimes \hilb_{\E} \otimes \hilb_{\E'}$
(formally this follows from the property of strong subadditivity of the von
Neumann entropy). In particular, we use this to replace $\rho_{\E}$ and
$\rho'_{\E}$ in \eqref{eq:cq_ae} with purifications $\ket{\Psi}$ and
$\ket{\Phi} \in \hilb_{\E} \otimes \hilb_{\E'}$. Applying the definition of
the conditional von Neumann entropy, we find
\begin{IEEEeqnarray}{rCl}
  \Hq(Z \mid \E) &\geq& \Hq(Z \mid \E\E') \IEEEnonumber \\
  &=& 1 - S \bigl[ \tfrac{1}{2} \bigl( \proj{\Psi}
  + \proj{\Phi} \bigr) \bigr] \IEEEnonumber \\
  &=& 1 - h \bigl[ \tfrac{1}{2}
  \bigl( 1 + \abs{\braket{\Psi}{\Phi}} \bigr) \bigr] \,,
\end{IEEEeqnarray}
where $S(\rho) = - \Tr[\rho \log(\rho)]$. Since by Uhlmann's theorem we can
choose $\ket{\Psi}$ and $\ket{\Phi}$ such that $F(\rho_{\E}, \rho'_{\E}) =
\abs{\braket{\Psi}{\Phi}}$, we find
\begin{equation}
  \label{eq:ent_fidel}
  \Hq(Z \mid \E) \geq 1 - h \bigl( \tfrac{1}{2}
   + \tfrac{1}{2} F(\rho_{\E}, \rho'_{\E}) \bigr) \,,
\end{equation}
which completes the derivation of \eqref{eq:keyrate_fidel}. Note that
\eqref{eq:ent_fidel} is equivalent to an analogous bound on the Holevo
quantity derived in \cite{ref:rfz2010}.

In the proof given here, we have made the tacit assumption that Alice emits
the $z$ states $\ket{\alpha}$ and $\ket{\alpha'}$ with equal probability. If
we wish to account for the possibility of a biased source, as was considered
in \cite{ref:mls2010} for instance, then it is straightforward to adapt the
derivation of \eqref{eq:keyrate_fidel}. In particular, if the two $z$-basis
states are emitted with probabilities $p = (1 + \varepsilon)/2$ and $q = (1 -
\varepsilon)/2$ respectively, \eqref{eq:ent_fidel} generalises to
\begin{equation}
  \label{eq:ent_fidel_biased}
  \Hq(Z \mid \E) \geq h(p)
  - h \Bigl[ \tfrac{1}{2} + \tfrac{1}{2} \sqrt{\varepsilon^{2} + (1 -
    \varepsilon^{2}) F(\rho_{\E}, \rho'_{\E})^{2}} \Bigr] \,.
\end{equation}
Our results will hold essentially unmodified if there is a bias in the
$x$ basis, provided the observed coincidence statistics are used to correctly
bound the trace distance $D(\sigma_{\B}, \sigma'_{\B})$. Here we have
departed from the approach used by the authors of
\cite{ref:gl2004,ref:k2009,ref:mls2010}, where any bias in the source was
included in their analogue of our source characterisation
\eqref{eq:theta_def}. This way we obtain a generally better result. For
instance, for an otherwise ideal BB84 implementation and in the absence of
errors, we obtain the best possible keyrate of $r = h(p)$.

\subsection{A note on non-asymptotic security\label{sec:finite_key}}

As alluded to above, the Devetak-Winter bound, and consequently the bound
\eqref{eq:keyrate_fidel} on the keyrate that we will use in the remainder of
this \work{}, is only guaranteed to hold in the asymptotic limit of an
infinitely long key. In the realistic finite-key case, a finite-key result
can be derived based on a ``one shot'' version of the Devetak-Winter rate
\cite{ref:tr2011} where the main problem in obtaining a security proof is
reduced to lower bounding the conditional min-entropy $H_{\mathrm{min}}(Z
\mid \E)$ rather than the von Neumann entropy $\Hq(Z \mid \E)$. To obtain a
result that converges asymptotically to the keyrate given by the
Devetak-Winter bound, the min-entropy should ideally be evaluated with a
smoothing parameter, which can roughly be interpreted as a small additional
probability of failure traded in exchange for a higher keyrate, and which can
be made arbitrarily small in the asymptotic limit. We refer to
\cite{ref:t2012,ref:tl2012} for a discussion of the details.

We will leave as a problem for future work how the smooth min-entropy can be
bounded in such a way as to recover \eqref{eq:ent_fidel} asymptotically. We
note only here that the min-entropy (without smoothing) evaluated on the
state \eqref{eq:cq_ae} has an exact expression in terms of the trace
distance:
\begin{equation}
  \label{eq:hmin_trdist}
  H_{\mathrm{min}}(Z \mid \E) = 1
  - \log \bigl( 1 + D(\rho_{\E}, \rho'_{\E}) \bigr) \,.
\end{equation}
This is a special case of the min-entropy's expression in terms of the
guessing probability \cite{ref:krs2009}; we also gave a simple derivation in
\cite{ref:wlp2013}. Using that $D(\rho_{\E}, \rho'_{\E}) \leq \sqrt{1 -
  F(\rho_{\E}, \rho'_{\E})^{2}}$, we can alternatively bound the min-entropy
in terms of the fidelity. While the result would be less than optimal and
still limited to collective attacks, it is worth noting that using
\eqref{eq:hmin_trdist} in place of \eqref{eq:ent_fidel} would already allow
the obtention of finite-key bounds using the general approach explored in
this \work{}.

\section{Bounding Eve's distinguishing ability\label{sec:fidel_bounds}}

\subsection{Ideal BB84 source}

In the special case of an ideal BB84 source, characterised by $\theta =
\pi/2$, the general cloning bound \eqref{eq:fidel_theta_bound} reduces to
\begin{equation}
  \label{eq:fidel_zx_bound}
  F(\rho_{\E}, \rho'_{\E}) \geq D(\sigma_{\B}, \sigma'_{\B}) \,,
\end{equation}
and we recover the famous keyrate $r \geq 1 - h(\delta_{x}) - h(\delta_{z})$
due to Shor and Preskill \cite{ref:sp2000}. While the proof of
\eqref{eq:fidel_theta_bound} is not complicated, a direct proof of
\eqref{eq:fidel_zx_bound} is especially simple and will serve to introduce
our techniques.

In an ideal BB84 implementation, the source states form two mutually unbiased
orthogonal bases, i.e.\ satisfy
\begin{eqnarray}
  \braket{\alpha}{\alpha'} = \braket{\beta}{\beta'} = 0
\end{eqnarray}
and, in an appropriate phase convention,
\begin{IEEEeqnarray}{rCl}
  \ket{\alpha} &=& \tfrac{1}{\sqrt{2}} [ \ket{\beta} + \ket{\beta'} ] \,,
  \IEEEnonumber \\
  \ket{\alpha'} &=& \tfrac{1}{\sqrt{2}} [ \ket{\beta} - \ket{\beta'} ] \,.
\end{IEEEeqnarray}
Equivalently, one can identify $Z = \rho - \rho'$ and $X = \sigma - \sigma'$
with the Pauli $z$ and $x$ operators. With this notation, $D(\rho_{\B},
\rho'_{\B}) = \tfrac{1}{2} \trnorm{Z_{\B}}$ and $D(\sigma_{\B},
\sigma'_{\B}) = \tfrac{1}{2} \trnorm{X_{\B}}$.

We obtain \eqref{eq:fidel_zx_bound} as follows. Let $U_{\mathrm{B}}$ be the
(Hermitian) unitary operator acting on $\hilb_{\B}$ such that $\tfrac{1}{2}
\trnorm{X_{\B}} = \tfrac{1}{2} \Tr_{\B}[ U_{\B} X_{\B} ]$. Then,
\begin{IEEEeqnarray}{rCl}
  \label{eq:fidel_zx_proof}
  D(\sigma_{\B}, \sigma'_{\B}) &=& \tfrac{1}{2} \Tr_{\B}[ U_{\B} X_{\B} ]
  \IEEEnonumber \\
  &=& \tfrac{1}{2} \Tr \bigl[ (U_{\B} \otimes \id_{\E}) X \bigr]
  \IEEEnonumber \\
  &=& \tfrac{1}{2} \bigl( \bra{\alpha} U_{\B} \otimes \id_{\E} \ket{\alpha'}
  + \bra{\alpha'} U_{\B} \otimes \id_{\E} \ket{\alpha} \bigr)
  \IEEEnonumber \\
  &=& \re \bigl[ \bra{\alpha} U_{\B} \otimes \id_{\E} \ket{\alpha'} \bigr]
  \IEEEnonumber \\
  &\leq& \babs{\bra{\alpha} U_{\B} \otimes \id_{\E} \ket{\alpha'}} \,.
\end{IEEEeqnarray}
Now, $\ket{\alpha}$ and $\ket{\alpha'}$ are by definition purifications of
$\rho_{\E}$ and $\rho'_{\E}$, and since $U_{\B} \otimes \id_{\E}$ acts
nontrivially only on $\hilb_{\B}$, the state $U_{\B} \otimes \id_{\E}
\ket{\alpha'}$ is still a purification of $\rho'_{\E}$. Thus, by Uhlmann's
theorem, the last line of \eqref{eq:fidel_zx_proof} provides a lower bound on
the fidelity between $\rho_{\E}$ and $\rho'_{\E}$, which concludes our proof.

We conclude this discussion of the ideal BB84 scenario by noting that
\eqref{eq:fidel_zx_bound} can be seen as a strengthened version of a bound
originally derived in \cite{ref:fg1997}. Using that $D(\rho_{\E}, \rho'_{\E})
\leq \sqrt{1 - F(\rho_{\E}, \rho'_{\E})^{2}}$, \eqref{eq:fidel_zx_bound}
implies
\begin{equation}
  \label{eq:tr_zx_bound}
  D(\rho_{\E}, \rho'_{\E})^{2} + D(\sigma_{\B}, \sigma'_{\B})^{2} \leq 1 \,.
\end{equation}
Although not given in this form, Fuchs \etal{} effectively derived this
relation as an intermediate result in their security proof of the BB84
protocol against individual attacks (specifically, \eqref{eq:tr_zx_bound} is
essentially equivalent to equations~(23) and (24) in \cite{ref:fg1997}).

\subsection{Arbitrary source states\label{sec:arbitrary_source}}

We now derive \eqref{eq:fidel_theta_bound}. Our starting point is the
quantity
\begin{equation}
  \label{eq:basis_overlap}
  \Delta = \tfrac{1}{2\sqrt{2}} \bigl( \braket{\alpha}{\beta} +
  \braket{\alpha'}{\beta} + \braket{\alpha}{\beta'} -
  \braket{\alpha'}{\beta'} \bigr) \,.
\end{equation}
Note that \eqref{eq:basis_overlap} as expressed above is dependent on
physically irrelevant phase factors (e.g.\ $\ket{\alpha}$ and $\ee^{\ii \phi}
\ket{\alpha}$ denote the same state); to obtain the best result, one should
adopt the phase convention that maximises \eqref{eq:basis_overlap}, which can
then always be taken to be real.

Let $P$ and $Q$ be orthogonal projectors such that $P - Q = U$ and $P + Q =
\id$, where $U = U_{\B} \otimes \id_{\E}$ is the Hermitian unitary such that
$\tfrac{1}{2} \Tr[U (\sigma - \sigma')] = D(\sigma_{B}, \sigma'_{\B})$, as in
the proof of \eqref{eq:fidel_zx_bound}. For convenience, we also define the
(generally not normalised) states $\ket{\alpha_{\pm}} = (\ket{\alpha} \pm
\ket{\alpha'})/\sqrt{2}$. Then,
\begin{IEEEeqnarray}{rCl}
  2 \Delta &=& \bra{\alpha_{+}} (P + Q) \ket{\beta}
  + \bra{\alpha_{-}} (P + Q) \ket{\beta'} \IEEEnonumber \\
  &\leq& \abs{\bra{\alpha_{+}} P \ket{\beta}}
  + \abs{\bra{\alpha_{+}} Q \ket{\beta}}
  \IEEEnonumber \\
  &&+\: \abs{\bra{\alpha_{-}} P \ket{\beta'}} + \abs{\bra{\alpha_{-}} Q
    \ket{\beta'}} \,.
\end{IEEEeqnarray}
Applying the Cauchy-Schwarz inequality twice on the first and fourth terms on
the right-hand side,
\begin{IEEEeqnarray}{rCl}
  &&\abs{\bra{\alpha_{+}} P \ket{\beta}}
  + \abs{\bra{\alpha_{-}} Q \ket{\beta'}} \IEEEnonumber \\
  &\leq& \sqrt{\bra{\alpha_{+}} P \ket{\alpha_{+}}}
  \sqrt{\bra{\beta} P \ket{\beta}}
  + \sqrt{\bra{\alpha_{-}} Q \ket{\alpha_{-}}}
  \sqrt{\bra{\beta'} Q \ket{\beta'}} \IEEEnonumber \\
  &\leq& \sqrt{\bra{\alpha_{+}} P \ket{\alpha_{+}}
    + \bra{\alpha_{-}} Q \ket{\alpha_{-}}}
  \sqrt{\bra{\beta} P \ket{\beta} + \bra{\beta'} Q \ket{\beta'}}
  \IEEEnonumber \\
  &=& \sqrt{1 + \re[\bra{\alpha} U \ket{\alpha'}]} 
  \sqrt{1 + D(\sigma_{\B}, \sigma'_{\B})} \,.
\end{IEEEeqnarray}
In a similar manner, $\abs{\bra{\alpha_{+}} Q \ket{\beta}} +
\abs{\bra{\alpha_{-}} P \ket{\beta'}}$ provides a lower bound on $\sqrt{1 -
  \re[\bra{\alpha} U \ket{\alpha'}]} \sqrt{1 - D(\sigma_{\B},
  \sigma'_{\B})}$. We thus obtain
\begin{IEEEeqnarray}{rCl}
  \label{eq:re_aa_delta_bound}
  \Delta &\leq& \sqrt{\tfrac{1}{2} + \tfrac{1}{2} \re[\bra{\alpha} U
    \ket{\alpha'}]} \sqrt{\tfrac{1}{2} + \tfrac{1}{2} D(\sigma_{\B},
    \sigma'_{\B})}
  \IEEEnonumber \\
  &&+ \sqrt{\tfrac{1}{2} - \tfrac{1}{2} \re[\bra{\alpha} U \ket{\alpha'}]}
  \sqrt{\tfrac{1}{2} - \tfrac{1}{2} D(\sigma_{\B}, \sigma'_{\B})} \,,
\end{IEEEeqnarray}
where $\re[\bra{\alpha} U \ket{\alpha'}]$ in turn provides a lower bound on
$F(\rho_{\E}, \rho'_{\E})$, as in the proof of
\eqref{eq:fidel_zx_bound}. Substituting $\Delta = \sqrt{1 +
  \abs{\sin(\theta)}} / \sqrt{2}$ and rearranging, we arrive at
\eqref{eq:fidel_theta_bound}.

If $\dim \hilb_{\A} = 2$ and the source states form two orthogonal bases,
then $\theta$ as defined by \eqref{eq:theta_def} coincides with the angle
separating the bases on the Bloch sphere. In this particular case, the source
is basis independent (i.e.\ $\tfrac{1}{2} \rho + \tfrac{1}{2} \rho' =
\tfrac{1}{2} \sigma + \tfrac{1}{2} \sigma'$), and we note that the keyrate
\eqref{eq:keyrate_theta} is an improvement over the keyrate predicted by the
uncertainty relation \cite{ref:bc2010,ref:tr2011}, which for comparison is
\begin{equation}
  \label{eq:keyrate_unc}
  r \geq 1 - \log(1 + \abs{\cos(\theta)})
  - h(\delta_{x}) - h(\delta_{z}) \,.
\end{equation}

A simple example demonstrates that the cloning bound
\eqref{eq:fidel_theta_bound} is tight. We set the source states to
\begin{IEEEeqnarray}{rCl}
  \ket{\alpha} &=&
  \cos \bigl( \tfrac{\gamma - \theta}{2} \bigr) \ket{0}_{\A}
  + \sin \bigl( \tfrac{\gamma - \theta}{2} \bigr) \ket{1}_{\A} \,,
  \IEEEnonumber \\
  \ket{\alpha'}
  &=& - \sin \bigl( \tfrac{\gamma - \theta}{2} \bigr) \ket{0}_{\A}
  + \cos \bigl( \tfrac{\gamma - \theta}{2} \bigr) \ket{1}_{\A} \,,
\end{IEEEeqnarray}
and
\begin{IEEEeqnarray}{rCl}
  \ket{\beta} &=&
  \cos \bigl( \tfrac{\gamma}{2} \bigr) \ket{0}_{\A}
  + \sin \bigl( \tfrac{\gamma}{2} \bigr) \ket{1}_{\A} \,,
  \IEEEnonumber \\
  \ket{\beta'}
  &=& - \sin \bigl( \tfrac{\gamma}{2} \bigr) \ket{0}_{\A}
  + \cos \bigl( \tfrac{\gamma}{2} \bigr) \ket{1}_{\A} \,,
\end{IEEEeqnarray}
where $\ket{0}_{\A}$ and $\ket{1}_{\A}$ are orthogonal. Following the cloning
operation
\begin{IEEEeqnarray}{rCl}
  \ket{0}_{\A} &\mapsto& \ket{0}_{\B} \ket{0}_{\E} \IEEEnonumber \\
  \ket{1}_{\A} &\mapsto& \ket{1}_{\B} \ket{1}_{\E} \,,
\end{IEEEeqnarray}
we find $D(\sigma_{\B}, \sigma'_{\B}) = \abs{\cos(\gamma)}$ and $F(\rho_{\E},
\rho'_{\E}) = \abs{\sin(\theta - \gamma)}$. If we take, for instance, angles
satisfying $0 \leq \gamma \leq \theta \leq \pi/2$, then $F(\rho_{\E},
\rho'_{\E}) = f_{\theta} \bigl( D(\sigma_{\B}, \sigma'_{\B})
\bigr)$. Consequently, the keyrate \eqref{eq:keyrate_theta} is the highest
that can be derived using the particular approach we have followed so far --
i.e.\ where the conditional entropy is bounded via the fidelity using only
the $x$-basis error rate and given only the state characterisation defined in
\eqref{eq:theta_def}. We describe how better bounds on the fidelity and
conditional entropy can be derived in certain circumstances of interest in
the next sections.

\subsection{Qubit source states\label{sec:qubit_source}}

We now turn to a more detailed study of qubit sources. It will be convenient
to characterise such a source in terms of three parameters: an angle
$\varphi$ representing the angle between the two bases on the Bloch sphere,
and angles $\alpha$ and $\beta$ measuring the non-orthogonality of the states
constituting each basis and defined by $\abs{\sin(\alpha)} =
\abs{\braket{\alpha}{\alpha'}}$ and $\abs{\sin(\beta)} =
\abs{\braket{\beta}{\beta'}}$ respectively. We set
\begin{IEEEeqnarray}{rCl}
  \rho - \rho' &\propto& Z \,, \IEEEnonumber \\
  \sigma - \sigma &\propto& V \,,
\end{IEEEeqnarray}
with $V = \cos(\varphi) Z + \sin(\varphi) X$, and where we can identify $Z$
and $X$ with the Pauli $z$ and $x$ operators, respectively.

Via results from the preceding section, we already know that
\begin{equation}
  \label{eq:trx_trv}
  \tfrac{1}{2} \trnorm{X_{\B}}
  \geq f_{\varphi} \bigl( \tfrac{1}{2} \trnorm{V_{\B}} \bigr) \,,
\end{equation}
where the function $f_{\varphi}$ is as defined in \eqref{eq:def_f_theta},
except that we use the angle $\varphi$ between the bases in place of the
basis overlap angle $\theta$ defined in \eqref{eq:theta_def}. If
$\ket{\alpha}$ and $\ket{\alpha'}$ are orthogonal, i.e.\ if $\sin(\alpha) =
0$, then according to \eqref{eq:fidel_zx_bound} Eve's fidelity is simply
bounded by $F(\rho_{\E}, \rho'_{\E}) \geq \tfrac{1}{2} \trnorm{X_{\B}}$. A
natural conjecture one could consider is that this bound still holds if
$\ket{\alpha}$ and $\ket{\alpha'}$ are non-orthogonal. This is not the case
however: we found counter-examples numerically. Nor is this an artefact of
bounding the fidelity as an intermediate step in bounding the keyrate, as we
likewise found cases where $\Hq(Z \mid \E) \ngeq 1 - h \bigl( \tfrac{1}{2} +
\tfrac{1}{4} \trnorm{X_{\B}} \bigr)$ when minimising the conditional von
Neumann entropy directly.

A generally valid bound, then, will depend explicitly on the
non-orthogonality measured by the angle $\alpha$. Specifically, we find
\begin{equation}
  \label{eq:fidel_alpha}
  F(\rho_{\E}, \rho'_{\E})
  \geq g_{\alpha} \bigl( \tfrac{1}{2} \trnorm{X_{\B}} \bigr)
\end{equation}
with the function $g_{\alpha}$ defined by
\begin{equation}
  \label{g_alpha_def}
  g_{\alpha}(x) =
  \begin{cases}
    (1 + \abs{\sin(\alpha)}) x - \abs{\sin(\alpha)}
    &: x \geq \frac{2 \abs{\sin(\alpha)}}{1 + \abs{\sin(\alpha)}} \\
    \abs{\sin(\alpha)}
    &: x \leq \frac{2 \abs{\sin(\alpha)}}{1 + \abs{\sin(\alpha)}}    
  \end{cases}  \,.
\end{equation}
Combining \eqref{eq:trx_trv} and \eqref{eq:fidel_alpha}, we have
\begin{equation}
  \label{eq:fidel_alpha_theta}
  F(\rho_{\E}, \rho'_{\E}) \geq g_{\alpha} \circ f_{\varphi} \Bigl(
  \tfrac{D(\sigma_{\B}, \sigma'_{\B})}{\abs{\cos(\beta)}} \Bigr) \,,
\end{equation}
where we have used that $\sigma - \sigma' = \cos(\beta) V$. We see here that
explicitly introducing the non-orthogonality of the $x$-basis states can only
improve the keyrate bound. This is not surprising since the $x$-basis states
are only used for the purpose of testing the channel. Explicitly writing the
resulting keyrate,
\begin{equation}
  \label{eq:keyrate_alpha_theta}
  r \geq 1 - h \Bigl[ \tfrac{1}{2}
  + \tfrac{1}{2} g_{\alpha} \circ f_{\varphi} \Bigl(
  \tfrac{\abs{1 - 2 \delta_{x}}}{\abs{\cos(\beta)}} \Bigr) \Bigr]
  - h(\delta_{z}) \,.
\end{equation}

The remainder of this section will be devoted to the determination of the
function $g_{\alpha}$. First, we set
\begin{IEEEeqnarray}{rCl}
  \label{eq:z_states}
  \ket{\alpha} &=& \cos(\tfrac{\alpha}{2}) \ket{0}
  + \ee^{\ii \phi} \sin(\tfrac{\alpha}{2}) \ket{1} \,, \IEEEnonumber \\
  \ket{\alpha'} &=& \sin(\tfrac{\alpha}{2}) \ket{0}
  + \ee^{\ii \phi} \cos(\tfrac{\alpha}{2}) \ket{1} \,,
\end{IEEEeqnarray}
such that $\braket{\alpha}{\alpha'} = \sin(\alpha)$ and $\rho - \rho' =
\cos(\alpha) Z$. As in preceding sections, we define the Hermitian unitary $U
= U_{\B} \otimes \id_{\E}$ such that $\tfrac{1}{2} \trnorm{X_{\B}} =
\tfrac{1}{2} \Tr[U X] = \re[\bra{0} U \ket{1}]$. Without loss of generality,
we can take the quantity $\Gamma = \bra{0} U \ket{1}$ to be real (if
necessary, this can be achieved by absorbing its phase into the phase $\phi$
already present in \eqref{eq:z_states}). A lower bound is already given by
\begin{equation}
  \label{eq:fidel_aa_U}
  F(\rho_{\E}, \rho'_{\E}) \geq \babs{\bra{\alpha} U \ket{\alpha'}} \,,
\end{equation}
however the result this would produce is not optimal. In order to obtain a
better result, note that any unitary of the form $\tilde{U}_{\B} \otimes
\id_{\E}$ can be used in place of $U$ in \eqref{eq:fidel_aa_U}. In
particular, since $U$ is Hermitian, the family of operators
\begin{equation}
  U(\gamma) = \ii \sin(\gamma) \id + \cos(\gamma) U
\end{equation}
are unitary and satisfy this requirement. Evaluating $\bra{\alpha} U(\gamma)
\ket{\alpha'}$, we find
\begin{IEEEeqnarray}{rCl}
  \bra{\alpha} U(\gamma) \ket{\alpha'} &=& \tfrac{1}{2} \sin(\alpha)
  \bigl( \bra{0} U(\gamma) \ket{0} + \bra{1} U(\gamma) \ket{1} \bigr)
  \IEEEnonumber \\
  &&+\: \ee^{\ii \phi} \cos(\tfrac{\alpha}{2})^{2} \bra{0} U(\gamma) \ket{1}
  + \ee^{-\ii \phi} \sin(\tfrac{\alpha}{2})^{2} \bra{1} U(\gamma) \ket{0}
  \IEEEnonumber \\
  &=& \ii \sin(\gamma) \sin(\alpha) + \cos(\gamma) \sin(\alpha) K
  \IEEEnonumber \\
  &&+\: \cos(\gamma) \bigl( \cos(\phi)
  + \ii \cos(\alpha) \sin(\phi) \bigr) \Gamma \,,
\end{IEEEeqnarray}
where we have set $K = \tfrac{1}{2} (\bra{0} U \ket{0} + \bra{1} U
\ket{1})$. Squaring the last line,
\begin{IEEEeqnarray}{rCl}
  \label{eq:aUa_sq}
  \babs{\bra{\alpha} U(\gamma) \ket{\alpha'}}^{2} &=&
  \cos(\gamma)^{2} \bigl( \sin(\alpha) K + \cos(\phi) \Gamma \bigr)^{2}
  \IEEEnonumber \\
  &&+\: \bigl( \sin(\gamma) \sin(\alpha)
  + \cos(\gamma) \cos(\alpha) \sin(\phi) \Gamma \bigr)^{2} \,.
\end{IEEEeqnarray}
Our task now is to maximise this quantity over $\gamma$ and minimise over
possible values of $K$ and $\phi$. We begin by minimising over $K$ for fixed
$\gamma$ and $\phi$. First, note that $\abs{K} + \Gamma \leq 1$ (this can be
inferred by evaluating \eqref{eq:aUa_sq} for $\gamma = \phi = 0$ and $\alpha
= \pm \pi/2$ and noting that the result should never exceed 1). With this
constraint, we wish to minimise $Q = \abs{\sin(\alpha) K + \cos(\phi)
  \Gamma}$. If $\Gamma$ is sufficiently small, we are able to choose $K$ such
that this quantity is zero. Otherwise, we simply set $\abs{K} = 1 -
\Gamma$. Calling $Q_{*}$ the minimum possible value of $\abs{\sin(\alpha) K +
  \cos(\phi) \Gamma}$, we have
\begin{equation}
  Q_{*} = \begin{cases}
    \bigl(\abs{\sin(\alpha)} + \abs{\cos(\phi)} \bigr) \Gamma -
    \abs{\sin(\alpha)}
    &: \Gamma \geq \frac{\abs{\sin(\alpha)}}{
      \abs{\sin(\alpha)} + \abs{\cos(\phi)}} \\
    0 &: \Gamma \leq \frac{\abs{\sin(\alpha)}}{
      \abs{\sin(\alpha)} + \abs{\cos(\phi)}} \,,
  \end{cases}
\end{equation}

The next step is the maximisation over $\gamma$. Since $Q_{*}$ has no
dependence on $\gamma$, this is straightforward. For the optimal value
$\gamma_{*}$ of $\gamma$, we find
\begin{IEEEeqnarray}{rCrl}
  \label{eq:aUa_sq_gmin}
  \babs{\bra{\alpha} U(\gamma_{*}) \ket{\alpha'}}^{2}
  &=& \IEEEeqnarraymulticol{2}{l}{
    \tfrac{1}{2} \bigl( Q\subsup{*}{2}
    + \cos(\alpha)^{2} \sin(\phi)^{2} \Gamma^{2}  + \sin(\alpha)^{2} \bigr)}
  \IEEEnonumber \\
  &&+\: \tfrac{1}{2} \Bigl\{ \bigl( Q\subsup{*}{2}
  + \cos(\alpha)^{2} \sin(\phi)^{2} \Gamma^{2}
  - \sin(\alpha)^{2} \bigr)^{2} & \IEEEnonumber \\
  &&+\: 4 \sin(\alpha)^{2} \cos(\alpha)^{2}
  \sin(\phi)^{2} \Gamma^{2} & \Bigr\}^{1/2} \,.
\end{IEEEeqnarray}

Finally, we minimise over $\phi$. For values of $\phi$ such that
$\abs{\cos(\phi)} \leq \abs{\sin(\alpha)} \frac{1 - \Gamma}{\Gamma}$, we set
$Q_{*} = 0$, and \eqref{eq:aUa_sq_gmin} simplifies to
\begin{equation}
  \label{eq:aUa_Qis0}
  \babs{\bra{\alpha} U(\gamma_{*}) \ket{\alpha'}}^{2}
  = \sin(\alpha)^{2} + \cos(\alpha)^{2} \sin(\phi)^{2} \Gamma^{2} \,.
\end{equation}
If $\Gamma \leq \frac{\abs{\sin(\alpha)}}{1 + \abs{\sin(\alpha)}}$, then
$Q_{*} = 0$ regardless of $\phi$, and the minimum of \eqref{eq:aUa_Qis0}
is
\begin{equation}
  \babs{\bra{\alpha} U(\gamma_{*}) \ket{\alpha'}} = \abs{\sin(\alpha)} \,.
\end{equation}
Otherwise, the minimum of \eqref{eq:aUa_sq} for values of $\phi$ where $Q_{*}
= 0$ is found by using the minimum allowed value of $\abs{\sin(\phi)}$ in
\eqref{eq:aUa_Qis0}, which simplifies to
\begin{equation}
  \label{eq:aUa_Qis0_min}
  \babs{\bra{\alpha} U(\gamma_{*}) \ket{\alpha'}}
  = \sin(\alpha)^{2} + \cos(\alpha)^{2} \Gamma \,.
\end{equation}

If $\Gamma > \frac{\abs{\sin(\alpha)}}{1 + \abs{\sin(\alpha)}}$, then we must
separately consider the range of possible values of $\phi$ for which
$\abs{\cos(\phi)} > \abs{\sin(\alpha)}\frac{1 - \Gamma}{\Gamma}$, where
$Q_{*} = (\abs{\sin(\alpha)} + \abs{\cos(\phi)}) - \abs{\sin(\alpha)}$. We
have
\begin{equation}
  Q\subsup{*}{2} + \cos(\alpha)^{2} \sin(\phi)^{2} \Gamma^{2}
  = Y \Gamma (Y \Gamma - 2) + 2 \cos(\alpha)^{2} \Gamma + \sin(\alpha)^{2} \,,
\end{equation}
where we have set $Y = 1 + \abs{\sin(\alpha)} \abs{\cos(\phi)}$. Then,
\begin{IEEEeqnarray}{rCll}
  \label{eq:aUa_sq_p}
  \babs{\bra{\alpha} U(\gamma_{*}) \ket{\alpha'}}^{2}
  &=& \IEEEeqnarraymulticol{2}{l}{\sin(\alpha)^{2} + \cos(\alpha)^{2} \Gamma
    + \tfrac{1}{2} Y \Gamma (Y \Gamma - 2)} \IEEEnonumber \\
  &&+\: \tfrac{1}{2} \Gamma \Bigl\{ & \bigl( Y (Y \Gamma - 2)
  + 2 \cos(\alpha)^{2} \bigr)^{2} \IEEEnonumber \\
  &&&+\: 4 \sin(\alpha)^{2} \cos(\alpha)^{2} \sin(\phi)^{2} \Bigr\}^{1/2} \,,
\end{IEEEeqnarray}
which we can simplify down to
\begin{IEEEeqnarray}{rCl}
  \babs{\bra{\alpha} U(\gamma_{*}) \ket{\alpha'}}^{2}
  &=& \sin(\alpha)^{2} + \cos(\alpha)^{2} \Gamma
  + \tfrac{1}{2} Y \Gamma (Y \Gamma - 2) \IEEEnonumber \\
  &&+\: \tfrac{1}{2} Y \Gamma \sqrt{(Y \Gamma - 2)^{2}
    + 4\cos(\alpha)^{2}(\Gamma - 1)} \,.
\end{IEEEeqnarray}
This is a decreasing function in $Y$ and is therefore minimised by taking
$\abs{\cos(\phi)} = 1$, or $Y_{*} = 1 + \abs{\sin(\alpha)}$. We find
\begin{IEEEeqnarray}{rCl}
  \babs{\bra{\alpha} U(\gamma_{*}) \ket{\alpha'}}^{2}
  &=& \sin(\alpha)^{2} + \cos(\alpha)^{2} \Gamma \IEEEnonumber \\
  &&+\: \tfrac{1}{2} Y_{*} \Gamma \bigl[ Y_{*} \Gamma - 2
  + \babs{Y_{*} \Gamma - 2 \abs{\sin(\alpha)}} \bigr] \,.
\end{IEEEeqnarray}
If $\Gamma \geq \frac{2 \abs{\sin(\alpha)}}{1 + \abs{\sin(\alpha)}}$, then
the minimum we find is
\begin{equation}
  \label{eq:aUa_Qisnot0_min}
  \babs{\bra{\alpha} U(\gamma_{*}) \ket{\alpha'}}
  = \bigl( 1 + \abs{\sin(\alpha)} \bigr) \Gamma - \abs{\sin(\alpha)} \,.
\end{equation}
Otherwise, the minimum is simply $\abs{\sin(\alpha)}$. Since
\eqref{eq:aUa_Qisnot0_min} is always less than \eqref{eq:aUa_Qis0_min}, our
final result is
\begin{equation}
  \babs{\bra{\alpha} U(\gamma_{*}) \ket{\alpha'}}
  \geq \begin{cases}
    \bigl( 1 + \abs{\sin(\alpha)} \bigr) \Gamma - \abs{\sin(\alpha)}
    &: \Gamma \geq \frac{2 \abs{\sin(\alpha)}}{1 + \abs{\sin(\alpha)}} \\
    \abs{\sin(\alpha)}
    &: \Gamma \leq \frac{2 \abs{\sin(\alpha)}}{1 + \abs{\sin(\alpha)}}
  \end{cases} \,.
\end{equation}
Recalling that $F(\rho_{\E}, \rho'_{\E}) \geq \abs{\bra{\alpha} U(\gamma_{*})
  \ket{\alpha'}}$ and that we defined $\Gamma = \tfrac{1}{2}
\trnorm{X_{\B}}$, this concludes the proof of \eqref{eq:fidel_alpha}.

\subsection{Qubit source and detector\label{sec:dim2_mes}}

In most security analyses of the BB84 protocol accounting for device
alignment errors, including the results derived in
sections~\ref{sec:arbitrary_source} and \ref{sec:qubit_source}, only the
$x$-basis error rate is used to bound the information an eavesdropper could
have about the key. This leaves open the possibility that we could derive
better results if the $z$-basis error rate were also used for this
purpose. Here, we will illustrate how such a result can be derived in the
case where Alice's source emits qubit states and Bob's detector is assumed to
perform two-dimensional measurements.

As in section~\ref{sec:qubit_source}, we take $\rho - \rho' \propto Z$ and
$\sigma - \sigma' \propto V$, where $V = \cos(\varphi) Z + \sin(\varphi) X$,
and $Z$ and $X$ can be identified with the Pauli $z$ and $x$ operators. The
error rates provide lower bounds on $D(\rho_{\B}, \rho'_{\B}) =
\abs{\cos(\alpha)} \tfrac{1}{2} \trnorm{Z_{\B}}$ and $D(\sigma_{\B},
\sigma'_{\B}) = \abs{\cos(\beta)} \tfrac{1}{2} \trnorm{V_{\B}}$. The main
intuition for deriving an improved keyrate is that, if $\dim \hilb_{\B} = 2$,
the quantities $\tfrac{1}{2} \trnorm{Z_{\B}}$, $\tfrac{1}{2}
\trnorm{X_{\B}}$, and $\tfrac{1}{2} \trnorm{V_{\B}}$ are constrained in the
values they can take. Specifically, we will be able to show that
\begin{equation}
  \label{eq:zvx_bound}
  \sqrt{1 - \tfrac{1}{4}\norm{V_{\B}}\subsup{1}{2}} \geq
  \abs{\sin(\varphi)} \sqrt{1 - \tfrac{1}{4} \norm{X_{\B}}\subsup{1}{2}}
  - \abs{\cos(\varphi)} \sqrt{1 - \tfrac{1}{4} \norm{Z_{\B}}\subsup{1}{2}} \,.
\end{equation}
Since we have $F(\rho_{\E}, \rho'_{\E}) \geq g_{\alpha} \bigl( \tfrac{1}{2}
\trnorm{X_{\B}} \bigr)$ from \eqref{eq:fidel_alpha}, \eqref{eq:zvx_bound}
provides a second bound on the fidelity to complement
\eqref{eq:fidel_alpha_theta}. Our end result is similar in form to
\eqref{eq:fidel_alpha_theta}:
\begin{equation}
  \label{eq:fidel_alpha_theta_dim2}
  F(\rho_{\E}, \rho'_{\E}) \geq g_{\alpha} \circ f^{(2)}_{\varphi} \Bigl(
  \tfrac{D(\rho_{\B}, \rho'_{\B})}{\abs{\cos(\alpha)}},
  \tfrac{D(\sigma_{\B}, \sigma'_{\B})}{\abs{\cos(\beta)}} \Bigr) \,,
\end{equation}
with the function $f^{(2)}_{\varphi}$ defined piecewise by
\begin{equation}
  \label{eq:def_f2_theta}
  f^{(2)}_{\varphi}(z, v) = 
  \begin{cases}
    h_{\varphi}(z, v) &: q(z, v) \geq \abs{\cos(\varphi)} \\
    f_{\varphi}(v) &: q(z, v) \leq \abs{\cos(\varphi)}
  \end{cases} \,,
\end{equation}
in turn with $h_{\varphi}$ defined such that
\begin{equation}
  \abs{\sin(\varphi)} \sqrt{1 - h_{\varphi}(z, v)^{2}}
  = \sqrt{1 - v^{2}} + \abs{\cos(\varphi)} \sqrt{1 - z^{2}} \,,
\end{equation}
$f_{\varphi}$ as in \eqref{eq:def_f_theta}, and $q$ by
\begin{equation}
  q(z, v) = zv - \sqrt{(1 - z^{2})(1 - v^{2})} \,.
\end{equation}
The resulting keyrate is
\begin{equation}
  \label{eq:keyrate_alpha_theta_dim2}
  r \geq 1 - h \Bigl[ \tfrac{1}{2}
  + \tfrac{1}{2} g_{\alpha} \circ f^{(2)}_{\varphi} \Bigl(
  \tfrac{\abs{1 - 2 \delta_{z}}}{\abs{\cos(\alpha)}},
  \tfrac{\abs{1 - 2 \delta_{x}}}{\abs{\cos(\beta)}} \Bigr) \Bigr]
  - h(\delta_{z}) \,.
\end{equation}

We note that, provided $\sin(\varphi) \neq 0$, if $\tfrac{1}{2}
\trnorm{Z_{\B}} = 1$ and $\tfrac{1}{2} \trnorm{V_{\B}} = 1$ then
\eqref{eq:zvx_bound} implies $\tfrac{1}{2} \trnorm{X_{\B}} = 1$. Thus, except
in the pathological case where the $z$ and $x$ bases coincide, if $\delta_{z}
= (1 - \abs{\cos(\alpha)})/2$ and $\delta_{x} = (1 - \abs{\cos(\beta)})/2$
(the minimum possible error rates) we certify that Bob is in full control of
Alice's source space and we find $\Hq(Z \mid \E) = 1$ and $r \geq 1 -
h(\delta_{z})$ (i.e.\ the only reduction in the keyrate is due to error
correction).

We broke \eqref{eq:zvx_bound} numerically in test cases where $\dim
\hilb_{\B} > 2$. In this respect, we see that there is an advantage to be
gained if some characterisation of Bob's detector is introduced in addition
to Alice's source.

We now prove \eqref{eq:zvx_bound}. Our strategy will be to derive an upper
bound on $\tfrac{1}{2} \trnorm{V_{\B}}$ given $\tfrac{1}{2} \trnorm{Z_{\B}}$
and $\tfrac{1}{2} \trnorm{X_{\B}}$, and then invert the resulting bound for
$\tfrac{1}{2} \trnorm{X_{\B}}$. Where $\hilb_{\B}$ is two dimensional, this
is especially easy as all three operators are expressible as combinations of
Pauli operators. We set $Z_{\B} = \vect{z} \cdot \sv$, $X_{\B} = \vect{x}
\cdot \sv$, and $V_{\B} = \vect{v} \cdot \sv$, with $\norm{\vect{z}},
\norm{\vect{x}}, \norm{\vect{v}} \leq 1$. For simplicity of notation we will
generally denote by e.g.\ $a$ the norm $\norm{\vect{a}}$ of a vector
$\vect{a}$. Because $a = \norm{\vect{a}} = \tfrac{1}{2} \trnorm{\vect{a}
  \cdot \sv}$, deriving the desired bound is reduced to bounding $v$ in terms
of $z$ and $x$. Using that $V_{\B} = \cos(\varphi) Z_{\B} + \sin(\varphi)
X_{\B}$, we have
\begin{IEEEeqnarray}{rCl}
  \label{eq:v_ito_zx}
  v &=& \tfrac{1}{2} \trnorm{V_{\B}} \IEEEnonumber \\
  &=& \tfrac{1}{2} \trnorm{\cos(\varphi) Z_{\B}
    + \sin(\varphi) X_{\B}} \IEEEnonumber \\
  &=& \norm{\cos(\varphi) \vect{z} + \sin(\varphi) \vect{x}} \,.
\end{IEEEeqnarray}
Squaring this and developing,
\begin{equation}
  \label{eq:v2_ito_zx}
  v^{2} = \tfrac{1}{2} (z^{2} + x^{2})
  + \tfrac{1}{2} \cos(2 \varphi) (z^{2} - x^{2})
  + \sin(2 \varphi) \vect{x} \cdot \vect{z} \,.
\end{equation}

Now, while we are working with a \emph{given} value of $\varphi$, we can
arrive at a nontrivial bound on $v$ by noting that $\tfrac{1}{2}
\trnorm{\cos(\varphi) Z_{\B} + \sin(\varphi) X_{\B}}$ must be less than 1 for
\emph{all} values of $\varphi$. The maximum value of \eqref{eq:v2_ito_zx}, if
$\varphi$ is allowed to take any value, is given by
\begin{equation}
  (v_{\mathrm{max}})^{2} = \tfrac{1}{2} (z^{2} + x^{2})
  + \sqrt{\tfrac{1}{4} (z^{2} - x^{2})^{2} + (\vect{x} \cdot \vect{z})^{2}} \,.
\end{equation}
Requiring that the right-hand side is less than 1, we find
\begin{equation}
  (2 - z^{2} - x^{2})^{2} \geq (z^{2} - x^{2})^{2}
  + 4 (\vect{x} \cdot \vect{z})^{2} \,,
\end{equation}
which simplifies to
\begin{equation}
  \abs{\vect{x} \cdot \vect{z}} \leq \sqrt{(1 - z^{2}) (1 - x^{2})} \,.
\end{equation}
From this, we deduce that
\begin{equation}
  v^{2} \leq \tfrac{1}{2} (z^{2} + x^{2})
  + \tfrac{1}{2} \cos(2\varphi) (z^{2} - x^{2}) 
  + \abs{\sin(2 \varphi)} \sqrt{(1 - z^{2}) (1 - x^{2})} \,,
\end{equation}
or equivalently,
\begin{equation}
  \sqrt{1 - v^{2}} \geq \babs{\abs{\cos(\varphi)} \sqrt{1 - z^{2}} -
  \abs{\sin(\varphi)} \sqrt{1 - x^{2}}} \,.
\end{equation}
Invoking the triangle inequality on \eqref{eq:v_ito_zx}, we also trivially
have
\begin{equation}
  v \leq \abs{\cos(\varphi)} z + \abs{\sin(\varphi)} x
\end{equation}
and
\begin{equation}
  v \geq \babs{\abs{\cos(\varphi)} z - \abs{\sin(\varphi)} x} \,.
\end{equation}
Inverting these, we find two lower bounds on $x$:
\begin{equation}
  \label{eq:rnx_z_v}
  \abs{\sin(\varphi)} \sqrt{1 - x^{2}} \leq \sqrt{1 - v^{2}}
  + \abs{\cos(\varphi)} \sqrt{1 - z^{2}}
\end{equation}
and
\begin{equation}
  \label{eq:x_z_v}
  \abs{\sin(\varphi)} x \geq \babs{v - \abs{\cos(\varphi)} z} \,,
\end{equation}
and the bound on $x$ we are seeking is simply whichever is the stronger of
the two, which depends on $z$ and $v$. Specifically, we should use
\eqref{eq:rnx_z_v} if $zv - \sqrt{(1 - z^{2})(1 - v^{2})} \geq
\abs{\cos(\varphi)}$, and \eqref{eq:x_z_v} otherwise.

Equation~\eqref{eq:rnx_z_v} is just a rearrangement of \eqref{eq:zvx_bound},
and for sufficiently good bounds on $z$ and $v$ provides a better bound on
$F(\rho_{\E}, \rho'_{\E})$ than
\eqref{eq:fidel_alpha}. Equation~\eqref{eq:x_z_v} by contrast proves to be of
little interest as in practice we have only lower bounds $z \geq z_{0}$ and
$v \geq v_{0}$ on $z$ and $v$. Because \eqref{eq:x_z_v} is always a
decreasing function of either $z$ or $v$, we cannot safely substitute lower
bounds $z_{0}$ and $v_{0}$ in their place.

For the sake of completeness, we show here that \eqref{eq:rnx_z_v} and
\eqref{eq:x_z_v} together are no more useful than \eqref{eq:rnx_z_v} and
\eqref{eq:fidel_alpha} in this respect. First, note that we certify nothing
if $v_{0} \leq \abs{\cos(\varphi)}$, as this permits $v =
\abs{\cos(\varphi)}$ and $z = 1$, for which both \eqref{eq:rnx_z_v} and
\eqref{eq:x_z_v} reduce to $x \geq 0$. If $v_{0} > \abs{\cos(\varphi)}$, then
\eqref{eq:x_z_v} is minimised by maximising $z$ (i.e.\ setting $z = 1$) and
minimising $v$ (i.e.\ setting $v = v_{0}$). The lower bound on $x$ implied by
\eqref{eq:rnx_z_v} is minimised by minimising both $z$ and $v$. In both cases
we can safely substitute $v = v_{0}$. If $z_{0}$ is such that $z_{0} v_{0} -
\sqrt{(1 - z\subsup{0}{2})(1 - v\subsup{0}{2})} \geq \abs{\cos(\varphi)}$,
then we simply use $z = z_{0}$ in \eqref{eq:rnx_z_v}. Otherwise, we use $z =
\abs{\cos(\varphi)} v_{0} + \abs{\sin(\varphi)} \sqrt{1 - v\subsup{0}{2}}$ in
either \eqref{eq:rnx_z_v} or \eqref{eq:x_z_v}, which both reduce to $x \geq
\abs{\sin(\varphi)} v_{0} - \abs{\cos(\varphi)} \sqrt{1 - v\subsup{0}{2}}$.

In summary, given lower bounds $z_{0}$ and $v_{0}$ on $z = \tfrac{1}{2}
\trnorm{Z_{\B}}$ and $v = \trnorm{V_{\B}}$ respectively, we have shown that
\begin{equation}
  x = \tfrac{1}{2} \trnorm{X_{\B}} \geq f^{(2)}_{\varphi}(z_{0}, v_{0}) \,,
\end{equation}
with the function $f^{(2)}_{\varphi}$ defined in \eqref{eq:def_f2_theta}
above. This concludes the derivation of \eqref{eq:fidel_alpha_theta_dim2}.

\section{Conclusion\label{sec:conclusion}}

We have shown that it is possible to relate the security of the
prepare-and-measure BB84 quantum key distribution protocol in a particularly
simple way to the inherent limits of cloning imposed by quantum physics. In
this respect, our approach is similar to that explored in
\cite{ref:bc2010,ref:tr2011} in that we relate security as closely as
possible to a generic characterisation of the structure of quantum physics,
with our cloning bounds such as \eqref{eq:fidel_theta_bound} playing a role
analogous to the uncertainty relations of
\cite{ref:bc2010,ref:tr2011}. Notably, in a departure from recent security
analyses, we see that a much more direct treatment of the prepare-and-measure
scenario is possible; recasting BB84 in its entanglement-based form is seen
not to be a necessary step prior to the analysis.

Our results hold with Bob's device left largely
uncharacterised. Specifically, our results hold wherever the Helstrom bound
does, which is at least wherever Bob's measurements are separable. We note
this is in line with a trend in recent security analyses featuring automatic
``one-sided device independence'' \cite{ref:bc2012}. (See also
\cite{ref:tf2013} for a recent security proof of BB84 that is completely
device independent on Bob's side.)

Our security proof, which is based on the Devetak-Winter bound, holds for
collective attacks, in which the eavesdropper is assumed to attack each qubit
unitarily individually and identically, but may delay her measurements until
after the classical post-processing has been applied. In the case of a BB84
implementation that is basis independent and where Bob's measurement devices
are memoryless, security against collective attacks implies security against
coherent attacks, as follows for instance from the post-selection technique
of \cite{ref:ckr2009}. The status of coherent attacks in more general
situations is still largely unknown, and we leave as an open question how our
results extend in this case.

We have also restricted our analysis to the asymptotic regime (though see
section~\ref{sec:finite_key}). To derive a security proof in the case of
finite statistics, it would be interesting to understand how our approach can
be adapted to bound the smooth min-entropy in a manner that recovers
\eqref{eq:ent_fidel} and cloning bounds such as \eqref{eq:fidel_theta_bound}
asymptotically. Additionally, as the main purpose of this \work{} is to
introduce an alternative approach to understanding QKD security, we have not
attempted to account for all possible issues that should be accounted for in
a truly unconditional security proof of practical QKD. In particular, for
simplicity we have assumed a lossless channel and perfect detector
efficiency. While this restriction can in principle be removed if Bob adopts
the convention of assigning outputs to non-detection events, in practice the
resulting effective error rates would be too high to certify the security of
a QKD scheme implemented with contemporary technology, and a better treatment
of losses and detector inefficiencies seems desirable. We refer to
\cite{ref:mls2010} for a possible approach treating non-detection as a third
possible outcome.

Finally, following the recent trend of partially or fully device-independent
approaches to QKD \cite{ref:ab2007,ref:pb2011,ref:lcq2012}, let us remark
that our security proof of BB84 depends, as in \cite{ref:mls2010}, on a
\emph{single} parameter $\theta$ characterising the quantum devices. Provided
that our techniques generalise to coherent attacks, this parameter would
represent the only feature that has to be trusted in an otherwise
device-independent implementation of BB84. It would be interesting then to
understand how this parameter could be ``self tested'' under some realistic
assumptions, hopefully then elevating the prepare-and-measure BB84 protocol
to a status competitive with existing partially device-independent approaches
to QKD.

\section*{Acknowledgements}
\addcontentsline{toc}{section}{Acknowledgements}

The author thanks Stefano Pironio for useful discussion and very helpful
suggestions and guidance during the preparation of this manuscript, and
acknowledges support from the Belgian Fonds pour la Formation \`a la
Recherche dans l'Industrie et dans l'Agriculture (F.R.I.A.), the EU projects
Q-Essence and QCS, the CHIST-ERA DIQIP project, the Interuniversity
Attraction Poles Photonics@be Programme (Belgian Science Policy), and the
FRS-FNRS under project DIQIP.

\addcontentsline{toc}{section}{References}
\bibliography{qkd_cloning}

\end{document}